\documentstyle[11pt,psfig,vsopsymp]{article}
\begin{document}


\markboth{\hfill Jones et al.}{Observations of a Sub-Parsec Accretion 
Disk \hfill}


\centerline{\LARGE\bf 
   VSOP Observations of a Sub-Parsec 
}\bigskip
\centerline{\LARGE\bf 
   Accretion Disk
}\bigskip

\centerline{\sc 
D.L.~Jones, A.E.~Wehrle, B.G.~Piner \& D.L.~Meier 
}\medskip

\centerline{\it
Jet Propulsion Laboratory, California Institute of Technology, USA
}

\begin{abstract} \noindent 
The physical conditions in the inner parsec of accretion disks 
believed to orbit the central black holes in active galactic 
nuclei can be probed by imaging the absorption of background
radio emission by ionized gas in the disk.  We report high
angular resolution observations of the nearby galaxy NGC 4261
which show evidence for free-free absorption by a thin, nearly
edge-on disk at several frequencies.  The angular width, and
probably the depth, of the absorption appears to increase with
decreasing frequency, as expected.  Because free-free absorption
is much larger at lower frequencies, the longest possible baselines 
are needed to provide adequate angular resolution; observing at
higher frequencies to improve resolution will not help. 
\end{abstract}


\sources{NGC4261, 3C270} 

\section{Introduction}

The nearby FR-I radio galaxy NGC 4261 (3C270) is a good candidate
for the detection of free-free absorption by ionized gas in an inner
accretion disk.  The galaxy is known to contain a central black hole
with a mass of $5 \times 10^{8} \ {\rm M}_{\odot}$, a nearly edge-on
nuclear disk of gas and dust with a diameter of $\approx 100$ pc, and
a large-scale symmetric radio structure which implies that the radio
axis is close to the plane of the sky.  At an assumed distance of 
40 Mpc, 1 milliarcsecond (mas) corresponds to 0.2 pc.  Previous VLBA 
observations of this galaxy revealed a parsec-scale radio jet and 
counterjet aligned with the kpc-scale jet (see Figure~\ref{Fig1}).
The opening angle of the jets is less than $20^{\circ}$ during the
first 0.2 pc and $< 5^{\circ}$ during the first 0.8 pc.  At 8.4 GHz  
we found evidence for a narrow gap in radio brightness at the base of
the parsec-scale counterjet, just east (left) of the brightest peak 
which we identified as the core based on its inverted spectrum between
1.6 and 8.4 GHz (see the left part of Figure~\ref{Fig2}, 
from Jones and Wehrle 1997).  We
tentatively identified this gap as the signature of free-free absorption
by a nearly-edge on inner disk with a width $<< 0.1$ pc and an average
electron density of $10^{3}-10^{8}\ {\rm cm}^{-3}$ over the 
inner 0.1 pc. 

\begin{center}
\begin{figure}
\vspace{46mm}
\includegraphics{VSOP-symp-fig1.ps}
\caption{VLBA image of NGC 4261 at 8.4 GHz.  The contours increase in steps 
of $\sqrt{2}$ starting at $\pm 0.75$\% of the peak, which is 99 mJy/beam. 
The restoring beam is $1.86 \times 0.79$ mas with major axis 
PA = $-1.3^{\circ}$.} \label{Fig1}
\end{figure}
\end{center}

\section{Observations}

We observed NGC 4261 at 1.6 and 4.9 GHz with HALCA and a ground array 
composed of 7 VLBA antennas plus Shanghai, Kashima, and the DSN 70-m
Tidbinbilla antennas at 1.6 GHz (22 June 1999) and 8 VLBA antennas
plus the phased VLA at 4.9 GHz (27 June 1999). 
During both epochs the VLBA antennas at St.~Croix
and Hancock were unavailable, as was the North Liberty antenna at 1.6 GHz. 
Data were recorded as two 16-MHz bandwidth channels with 
2-bit sampling by the Mark-III/VLBA systems and correlated at the VLBA
processor in Socorro.  Both channels were sensitive to left circular
polarization.

Fringe-fitting was carried out in AIPS after applying {\it a priori}
amplitude calibration.  For VLBA antennas we used continuously measured  
system temperatures, for the VLA we used measured ${\rm T}_{\rm A} /
{\rm T}_{\rm SYS}$ values with an assumed source flux density of 5 Jy, and 
for the remaining antennas we used typical gain and system temperature
values obtained from the VSOP web site.  Fringes were found to all
antennas at 1.6 GHz except HALCA, but the signal/noise ratio to 
Shanghai and Kashima was very low and these data were not used.  The
{\it a priori} amplitude calibration for Tidbinbilla was dramatically
incorrect for unknown reasons.  We calibrated Tidbinbilla by imaging
the compact structure of the source using VLBA data, then holding the
VLBA antenna gains fixed and allowing the Tidbinbilla gain to vary. 
This produced a good match in correlated flux density where the
projected VLBA and Tidbinbilla baselines overlap.  At 4.9 GHz fringes
were found to all antennas, including HALCA.  A similar correction
to the {\it a priori} amplitude calibration for HALCA and the phased 
VLA was applied. 

In both observations we found that averaging in frequency over both 
16-MHz channels in AIPS produced large, baseline-dependent amplitude
reductions even though the post-fringe-fit visibility phases were 
flat and continuous between channels.  Averaging over frequency within
each 16-MHz band separately fixed this problem.  Difmap was 
used for detailed data editing, self-calibration, and image
deconvolution.  Both 16-MHz bands were combined during imaging.

Imaging within Difmap used uniform weighting with the weight of HALCA
data increased by a factor of 500.  Several iterations of phase-only
self calibration, followed by amplitude self calibration iterations 
with decreasing time scales, resulted in good fits ($\chi^{2} \approx
1$) between the source model and the data. 

\section{Results}

\subsection{1.6 GHz}

Although our image at 1.6 GHz does not include data from HALCA, it 
does have more than twice the angular resolution of our previous 
1.6 GHz image (Jones and Wehrle 1997) due to the addition of Tidbinbilla.
The previous image showed a symmetric structure, with the jet and
counterjet extending west and east from the core.  No evidence
for absorption is seen in this image.  However, with higher resolution
we do detect a narrow gap in emission just east of the core, at the
base of the counterjet.  The width of the gap is less than 2 mas.

\subsection{4.9 GHz}

We detected fringes to HALCA at 4.9 GHz when the projected Earth-space
baselines were less than one Earth diameter.  The HALCA data fills 
in the (u,v) coverage hole between continental VLBA baselines and
those to Mauna Kea, and also increases the north-south resolution
by a factor of two.  Our 4.9 GHz image is shown in Figure~\ref{Fig2}.
Note that the gap in emission is again seen just east of the
peak.  A careful comparison of brightness along the radio axis
at 4.9 and 8.4 GHz shows that the gap is both deeper and wider
at 4.9 GHz, as expected from free-free absorption.  The region of
the gap has a very inverted spectrum, the brightest peak (core)
has a slightly less inverted spectrum, and the distant parts of
both the jet and counterjet have steep spectra.

\begin{center}
\begin{figure}
\vspace{57mm}
\includegraphics{VSOP-symp-fig2.ps}
\includegraphics{VSOP-symp-fig3.ps}
\caption{Grey-scale images of the nucleus of NGC 4261 at 8.4 GHz (left) and 
4.9 GHz (right), with identical fields of view.} \label{Fig2} 
\end{figure}
\end{center}

\section{Summary}

Our observations at 1.6 and 4.9 GHz appear to confirm the free-free
absorption explanation for the sub-parsec radio morphology in 
NGC 4261.  Measurements of the optical depth in the absorbed region
and the distance between the absorption and the core as a function
of frequency will allow the radial distribution of electron density
in the inner parsec of the disk to be determined.  

\acknowledgements

We gratefully acknowledge the VSOP Pro-ject, which is led by the
Japanese Institute of Space and Astronautical Science in cooperation
with many organizations and radio telescopes around the world.
This research was carried out at the Jet Propulsion Laboratory,
California Institute of Technology, under contract with the
U.S.~National Aeronautics and Space Administration.

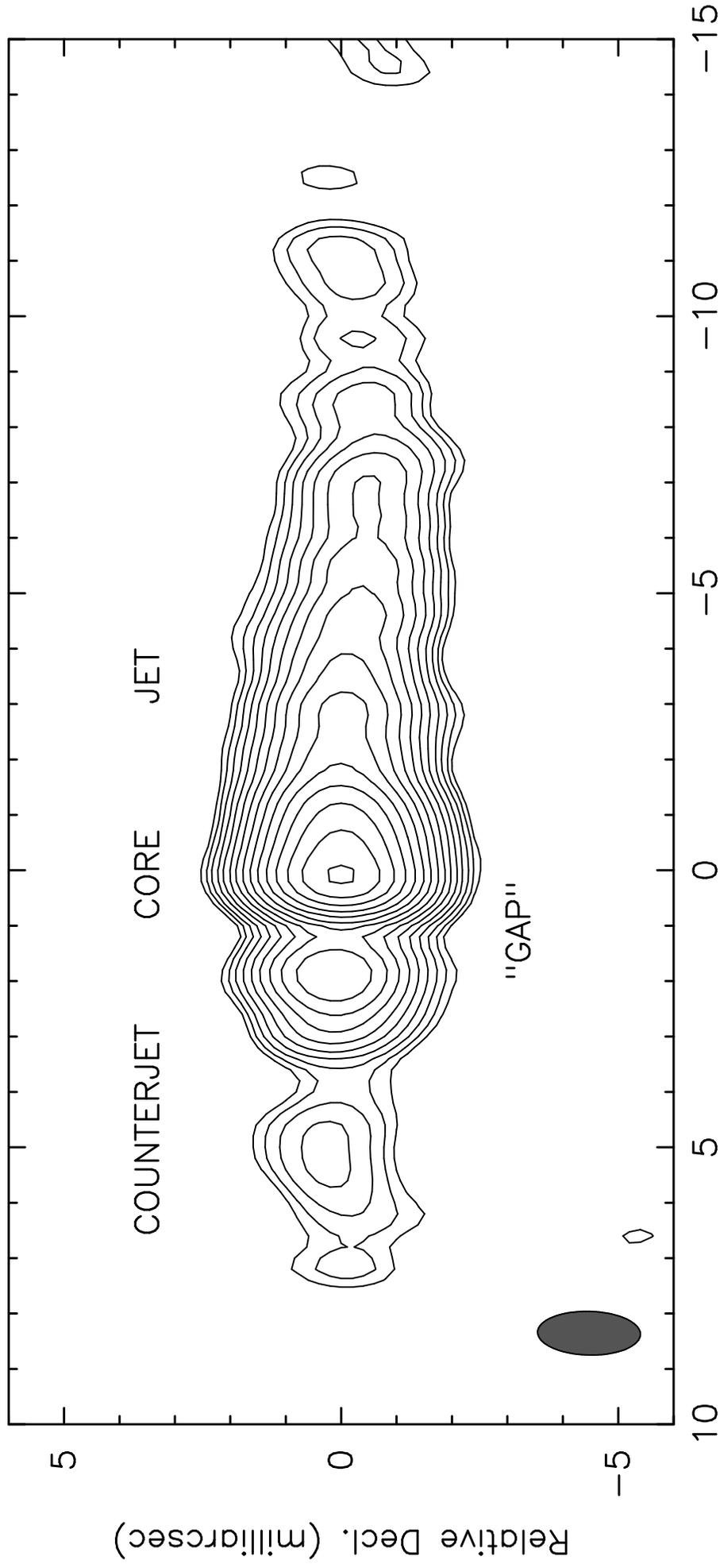
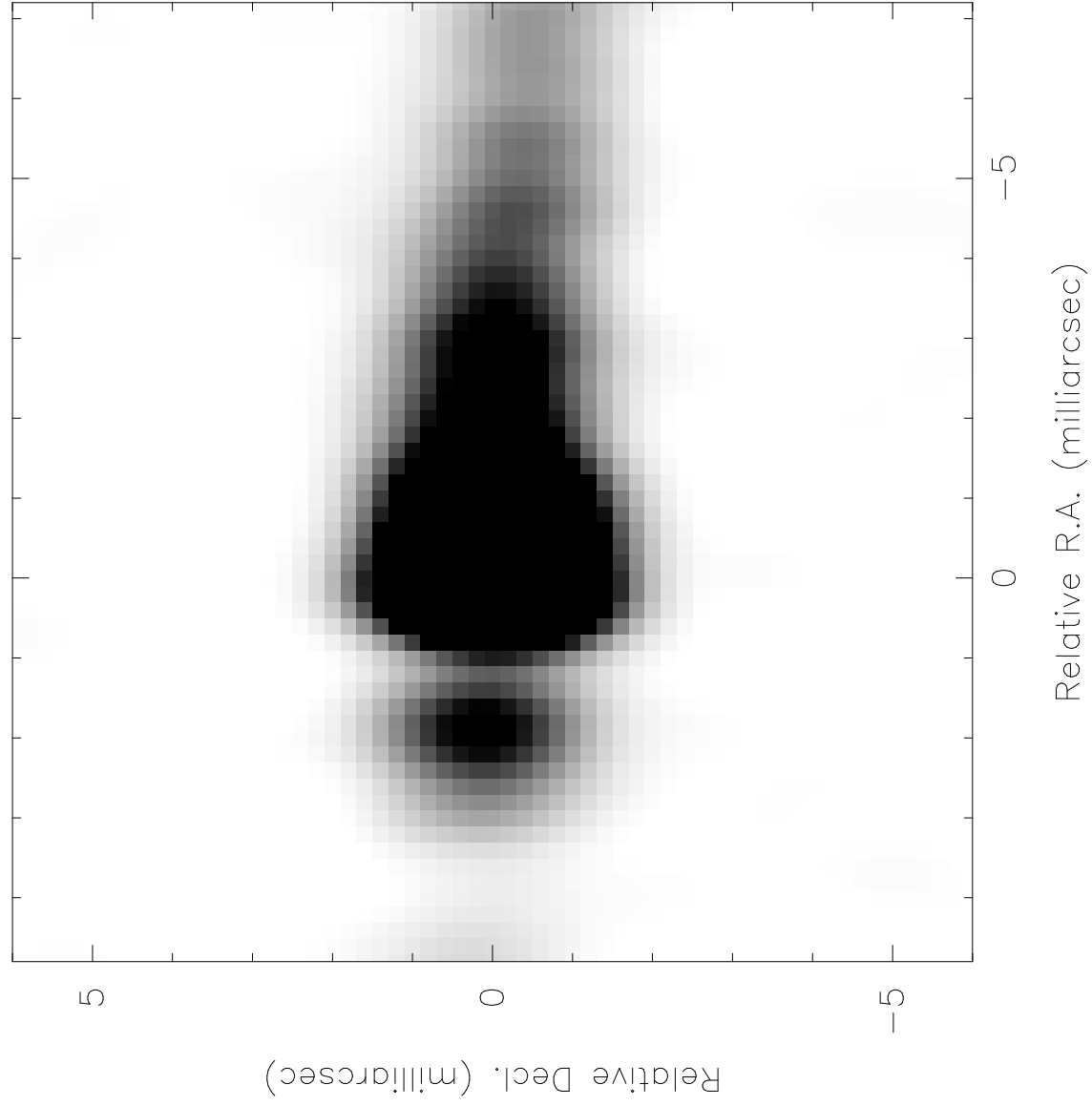
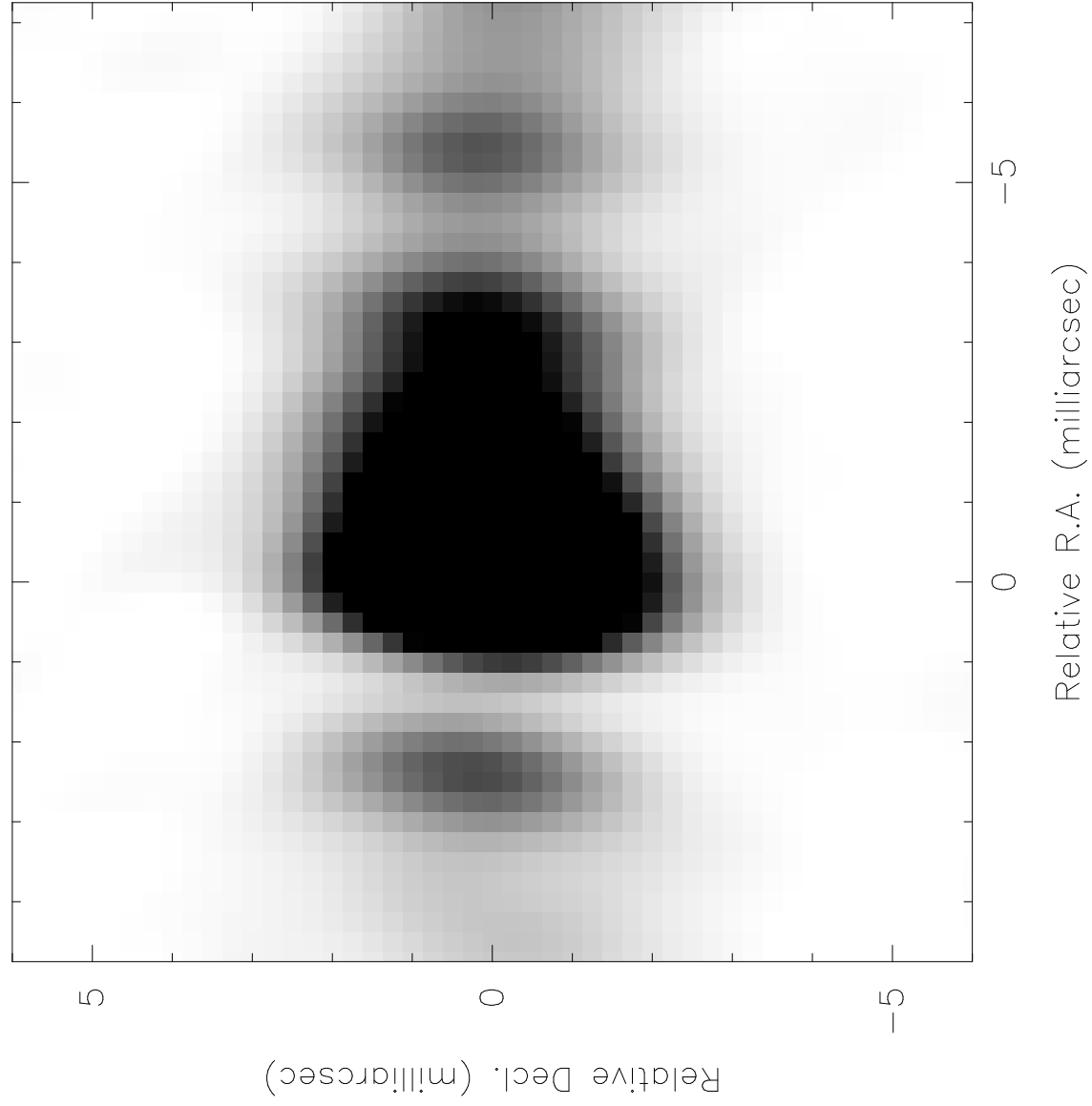
\end{document}